\documentclass{article} 
\usepackage[preprint]{colm2026_conference}

\usepackage{microtype}
\usepackage{hyperref}
\usepackage{url}
\usepackage{booktabs}

\usepackage{array}
\usepackage{booktabs}
\usepackage{ragged2e}

\newcolumntype{M}[1]{>{\centering\arraybackslash}m{#1}}
\newcolumntype{J}[1]{>{\centering\arraybackslash}m{#1}}

\definecolor{headerpurple}{RGB}{103, 80, 164}
\definecolor{rowpurple}{RGB}{245, 242, 252}

\definecolor{headerorange}{RGB}{230, 145, 56}
\definecolor{roworange}{RGB}{252, 244, 232}

\definecolor{rowwhite}{RGB}{255, 255, 255}

\usepackage[most]{tcolorbox}
\usepackage{xcolor}

\definecolor{rqBlueBg}{RGB}{235,242,250}

\usepackage[most]{tcolorbox}
\usepackage{xcolor}
\usepackage{soul}

\definecolor{rqBlueBg}{RGB}{235,242,250}
\definecolor{rqBlueText}{RGB}{35,55,90}
\definecolor{rqHighlight}{RGB}{255,245,180}
\definecolor{rowgray}{RGB}{248, 248, 248}

\tcbset{
  aibox/.style={
    width=\linewidth,
    top=5pt,
    bottom=3pt,
    left=4pt,
    right=4pt,
    colback=rowpurple,
    colframe=headerpurple,
    colbacktitle=headerpurple,
    coltitle=white,
    boxrule=0.6pt,
    enhanced,
    attach boxed title to top left={yshift=-2mm,xshift=3mm},
    boxed title style={
      boxrule=0pt,
      colframe=headerpurple,
      colback=headerpurple,
      sharp corners
    },
    fonttitle=\bfseries\small,
    title filled=true
  }
}

\newtcolorbox{augheuristic}[2][]{
  aibox,
  title={#2},
  #1
}

\usepackage{graphicx}
\usepackage{subcaption}
\usepackage{multirow}
\usepackage{graphicx} 
\usepackage{hyperref}

\usepackage[table]{xcolor}

\NewDocumentCommand{\heng}
{ mO{} }{\textcolor{red}{\textsuperscript{\textit{Heng}}\textsf{\textbf{\small[#1]}}}}

\NewDocumentCommand{\abbasi}
{ mO{} }{\textcolor{red}{\textsuperscript{\textit{abbasi}}\textsf{\textbf{\small[#1]}}}}

\NewDocumentCommand{\jiateng}
{ mO{} }{\textcolor{orange}{\textsuperscript{\textit{jiateng}}\textsf{\textbf{\small[#1]}}}}

\NewDocumentCommand{\qingyun}
{ mO{} }{\textcolor{blue}{\textsuperscript{\textit{Qingyun}}\textsf{\textbf{\small[#1]}}}}

\usepackage{lineno}

\definecolor{darkblue}{rgb}{0, 0, 0.5}
\hypersetup{colorlinks=true, citecolor=darkblue, linkcolor=darkblue, urlcolor=darkblue}

\usepackage{amsmath}
\usepackage{amssymb}
\usepackage{mathtools}
\usepackage{amsthm}

\usepackage{array}
\usepackage{booktabs} 
\usepackage{ragged2e} 

\usepackage[capitalize,noabbrev]{cleveref}
\usepackage{comment}

\theoremstyle{plain}

\theoremstyle{definition}

\theoremstyle{remark}

\title{Augmenting Interface Usability Heuristics\\ for Reliable Computer-Use Agents}

\author{
\noindent Jiateng Liu\textsuperscript{$1$}, Rushi Wang\textsuperscript{$1$}, Bingxuan Li\textsuperscript{$1$}, Kunlun Zhu\textsuperscript{$1$},\\
\bfseries\ Yifan Shen\textsuperscript{$1$}, Qingyun Wang\textsuperscript{$2$}, Ahmed Abbasi\textsuperscript{$3$}, Denghui Zhang\textsuperscript{$4$}, \bfseries\ Heng Ji\textsuperscript{$1$}, \vspace{4mm} \\
\textsuperscript{$1$}University of Illinois Urbana-Champaign,  
\textsuperscript{$2$}The College of William \& Mary, \\
\textsuperscript{$3$}University of Notre Dame, 
\textsuperscript{$4$}Stevens Institute of Technology
}

\begin{document}

\ifcolmsubmission
\linenumbers
\fi

\maketitle

\begin{abstract}
Recent advances have enabled general computer-use agents that interpret screens and execute grounded actions from human instructions, yet they still struggle to generalize to unseen and evolving interfaces. While improving agent capability remains important, agent-compatible interface design offers a complementary path by aligning interaction semantics with agents’ prior knowledge. In this paper, we revisit Nielsen’s 10 usability heuristics through the lens of computer-use agents, identifying which principles naturally transfer, where implicit design assumptions create agent-specific failures, and how safe additive augmentations can improve robustness without harming human usability. To evaluate these ideas, we introduce UI-Verse, a suite of controlled environments built around functionally similar interfaces with different applied heuristics. Experiments show that our augmented heuristics consistently improve task completion and modestly improve efficiency, with combined heuristics yielding further gains. Human studies further show that these designs preserve the original interaction workflow without observable usability regressions. Overall, our findings highlight interface design as a practical complementary avenue for improving the reliability and generalization of computer-use agents.
\end{abstract}

\section{Introduction}

As software interfaces become the front door to both everyday life and professional work, general computer-use agents (CUAs) are emerging as a new route to automation: systems that operate software directly, much like humans do~\citep{,agashe2024agentsopenagentic,wang2025opencua,hu2025osagentssurveymllmbased}. Powered by large multimodal models, they are able to interpret the screen and take grounded actions to handle tasks like online shopping, web browsing, replying to emails, and even help with  daily work~\citep{xie2024osworldbenchmarkingmultimodalagents,wang2024officebenchbenchmarkinglanguageagents,bonatti2024windowsagentarenaevaluating,pan2024webcanvasbenchmarkingwebagents,rawles2025androidworlddynamicbenchmarkingenvironment}. 

However, because real world interfaces are continually evolving and highly dynamic, today’s general CUAs still struggle to generalize to unseen UIs. In practice, they frequently rely on test time scaling and trial and error to navigate to new environments. They behave more like novices than professionals who operate through stable procedural knowledge \citep{abhyankar2025osworldhuman,liu2026osexpert}. This limitation is typically treated as a model capability problem. Yet generalization in interactive environments is not determined solely by the agent; it is also shaped by how interface states are represented, how actions are exposed, and how consistently interaction semantics are organized.

\begin{figure*}[t]
    \centering
    \includegraphics[width=1.0\columnwidth]{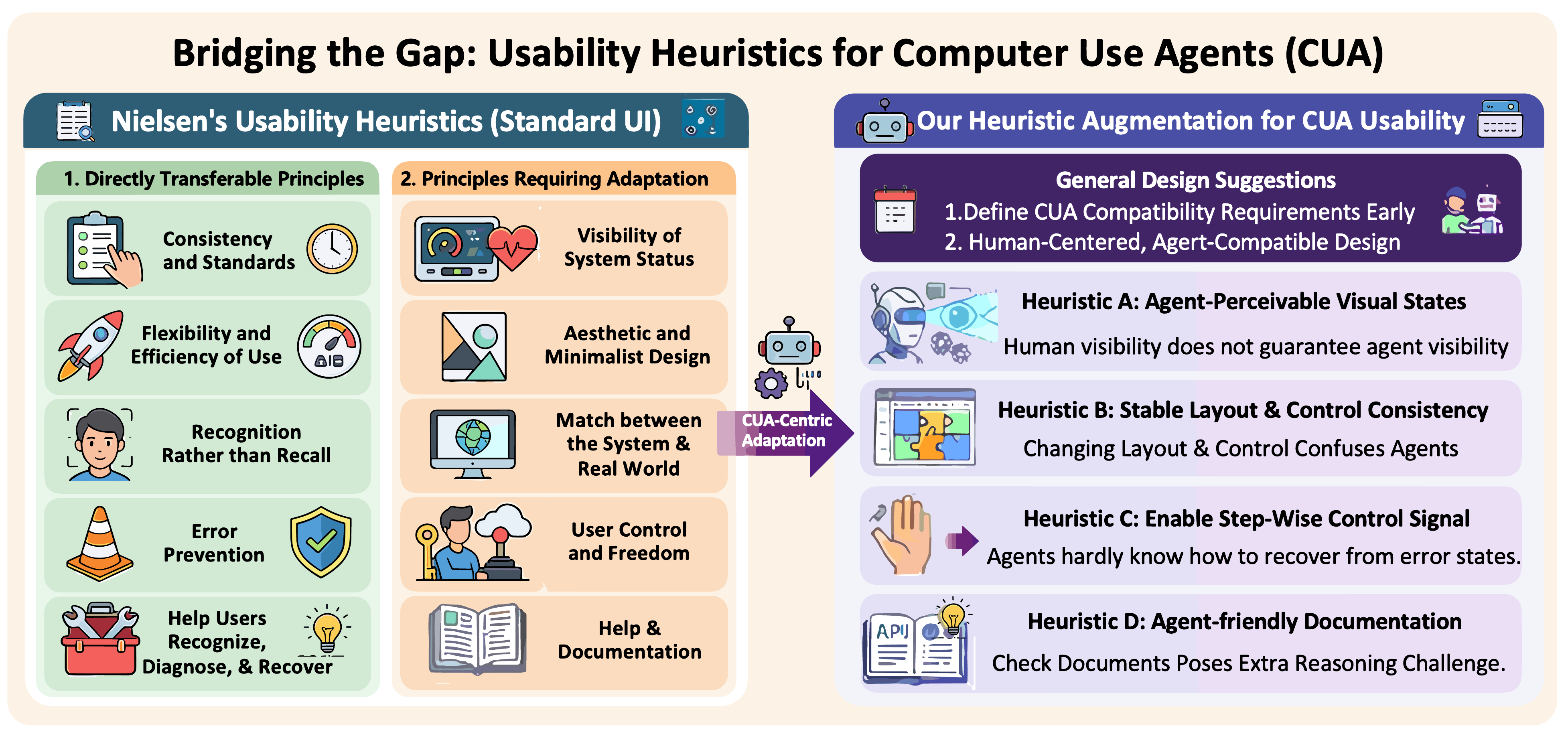}
    \caption{The left panel revisits Nielsen’s 10 classical usability heuristics~\cite{nielsen1994heuristics}, originally proposed for \emph{human} users, through the lens of general CUAs. Heuristics that benefit both humans and CUAs transfer directly (green), while those that may impair reliable CUA perception, reasoning, or grouding require adaptation (orange). The right panel presents our extension: two core design principles and four heuristics for designing interfaces that remain human-centered while better supporting reliable CUAs.}
    \label{fig:examples}
    \vspace{-6pt}
\end{figure*}

While improving agent architectures remains essential, we argue that interface design itself plays a critical and under-explored role in enabling reliable automation. Agent compatible interface design can substantially reduce adaptation burden by aligning system states and interaction structures with agents’ prior knowledge, thereby improving robustness and performance without modifying the underlying model. Motivated by this perspective, we revisit Nielsen’s ten usability heuristics~\citep{nielsen1990heuristic} for human centered interface design and systematically examine their implications for computer use agents’ perception, grounding, and control. As indicated in Figure~\ref{fig:examples}(Left), our analysis (Section \S ~\ref{sec: revisiting}) reveals that existing heuristics are not uniformly aligned with the needs of automation with CUAs. We therefore categorize them into two types. The first group consists of directly transferable principles, where stronger adherence consistently benefits both humans and agents. The second group includes principles that require CUA specific adaptations. These principles remain valuable for human usability, but when automation is an explicit objective, they should be augmented with additional design considerations to ensure stable agent perception, reduce reasoning burden, and support reliable action grounding.

Based on this distinction, we provide two levels of guidance for future interface design. First, at the broader design level, we offer two general suggestions, summarized in Section \S~\ref{principle: core}: (1) consider CUA compatibility early in the design process, and (2) adopt a human-centered yet agent-aware design perspective. Second, when supporting computer-use agent automation is an explicit goal, we introduce four agent-specific heuristics as structured augmentations: (A) make visual states readily perceivable to agents, (B) maintain stable layouts and consistent controls, (C) provide step-wise control signals, and (D) expose agent-friendly documentation and explicit skills. Section \S~\ref{principle: augment} presents the formal definitions, theoretical grounding, and concrete implementation examples for these heuristics.

To validate the impact of UI design on computer use and assess the effectiveness of our proposed heuristics for improving CUAs, we construct UI-Verse (Detailed in Appendix ~\ref{app:uiverse}), a suite of controlled digital environments. These environments provide similar functionalities and require CUAs to complete comparable task goals, while differing in the usability heuristics reflected in their UI designs. We evaluate current CUAs in terms of both task success rate and efficiency, and further analyze their interaction trajectories by attributing the root causes of failed interactions. Our results show that UI design alone can substantially affect CUA performance. Across environments and design variants, interfaces aligned with any of our proposed heuristics consistently yield more reliable agent behavior, higher task success rates, and sometimes efficiency gains. When these heuristics are implemented together, their benefits are complementary, leading to peak performance gains of up to 0.46 absolute gain on related tasks. Our error-type analysis further provides insight and helps explain the mechanisms behind the observed gains. Finally, we conduct comprehensive human studies to verify that these heuristic-based augmentations, while improving automation performance, do not necessarily compromise interface aesthetics, disrupt the original human workflow, or degrade usability for human users.

\textbf{Overall, our contributions are as follows:}

\begin{itemize}
    \item We identify agent-compatible interface design as an important but underexplored challenge for general CUAs, and argue that robust automationdepends not only on agent capability, but also on how interfaces expose states and actions.

    \item We propose a human-centered, agent-compatible interface design framework with two general design suggestions and four agent-specific heuristics that extend classical usability principles to better support agent perception, grounding, and control.

    \item We validate this framework through agent evaluations and human studies, showing that these heuristics consistently improve CUA reliability and efficiency while preserving human usability when designed appropriately.
\end{itemize}

\section{Related Work}

\paragraph{User-Centric Interface Design.}
User-centric interface design studies how interfaces help people perceive system state, act, and recover from errors. Foundational work introduced principles such as visibility, consistency, error prevention, and recovery support \citep{norman1983design,nielsen1990heuristic,nielsen1994enhancing}. Related work on direct manipulation and minimalist design further emphasized continuous visual representations, incremental and reversible actions, and immediately observable effects, making interfaces more predictable and easier to control \citep{kang2007minimalist,shneiderman1997direct}. These ideas remain influential in AI-mediated settings. Human-centered AI emphasizes preserving human agency and helping users interpret and guide intelligent systems \citep{capel2023human}, while later work extends classical design principles to generative AI applications, GUI-based machine learning tools, and AI-assisted design workflows \citep{weisz2024design,yamani2024establishing,lu2024ai}. However, these efforts still primarily optimize interfaces for \emph{human} users and do not systematically address settings where reliable operation by general UI-centered agents is also a core objective. Our work builds on this literature but shifts the focus toward human-centered, agent-compatible interface design.


\paragraph{General Computer-Use Agents.}
Recent work has substantially advanced general computer-use agents. Benchmarks such as OSWorld, OfficeBench, and Windows Agent Arena evaluate agent performance in realistic environments \citep{xie2024osworld,wang2024officebenchbenchmarkinglanguageagents,bonatti2024windows}. Most prior work, however, focuses on improving the agent itself. UI-TARS and OpenCUA develop native GUI agents with screenshot-based perception, action modeling, and deliberate reasoning \citep{qin2025uitarspioneeringautomatedgui,wang2025opencua}. Agent-S and CoAct further improve performance through stronger reasoning, memory, and Best-of-$N$ sampling \citep{agashe2024agentsopenagentic,song2025coact1}, while OSExpert shows that exploration-based learning can help agents acquire environment-specific skills \citep{liu2026osexpert}. However, existing work largely treats the UI as fixed and places the burden of adaptation on the agent. In contrast, we treat interface design as an optimization target, shifting part of the burden to the environment to better support reliable human-agent collaboration.

\section{From Classical Usability to Agent-Compatible Interface Design}
\label{sec:new heuristics}

Reliable automation with CUAs depends not only on agent capability, but also on whether interfaces expose states, actions, and transitions in ways that agents can consistently perceive, interpret, and execute. We therefore treat interface design not as a passive backdrop, but as an active factor shaping agent generalization. Our goal is not to replace classical human-centered usability principles, but to revisit and extend them for a setting in which interfaces are shared by both humans and agents, supporting effective collaboration while preserving human supervision. More broadly, we envision future digital workflows as human-centered and AI-assisted. To support this vision, we aim to design interfaces that agents can generalize to reliably while remaining understandable to humans and providing clear workflows for oversight. In this work, we proceed in three steps: first, we revisit Nielsen’s ten usability heuristics through the lens of computer-use agents (Section \S~\ref{sec: revisiting}); second, we offer general design suggestions for considering automation goals and agent compatibility early in the design process (Section \S~\ref{principle: core}); third, we operationalize these ideas into concrete CUA-specific heuristics that address agent perception, grounding, and control, enabling more robust generalization across environments (Section \S~\ref{principle: augment}).

\begin{table}[t]
\centering
\small
\setlength{\tabcolsep}{4pt}
\renewcommand{\arraystretch}{1.28}
\rowcolors{2}{roworange}{rowgray}
\resizebox{\columnwidth}{!}{
\begin{tabular}{M{2.7cm} J{7.0cm} J{5.0cm}}
\toprule
\rowcolor{headerorange}
\color{white}\textbf{Original Heuristic} &
\color{white}\textbf{Reasoning for Heuristic Adaptation} &
\color{white}\textbf{Example Failure Mode} \\
\midrule

\textbf{Visibility of System Status}
& Status visible to humans may be unreliable for CUAs, which perceive interfaces through discrete screenshots rather than continuous observation.
& The UI signals a state through transient or hover feedback, which the agent misses and misinterprets. \\

\midrule

\textbf{Aesthetic and Minimalist Design}
& Visual simplicity and aesthetic design may reduce explicit cues that CUAs rely on for grounding, even when the UI remains intuitive for humans.
& The UI uses icon-only controls or visually similar options, causing the agent to confuse and error. \\

\midrule

\textbf{Match Between the System and the Real World}
& Human-friendly metaphors and conventions may not always align well with CUA learnt priors when interface semantics are implied by real-world analogies rather than stated explicitly.
& The UI uses swiping to flip pages, following a physical book metaphor, but the agent fails to infer this convention and makes errors. \\

\midrule

\textbf{User Control and Freedom}
& Control and recovery mechanisms intuitive to humans may be inaccessible to CUAs, and multiple available workflows may confuse agents
& The agent fails to find a recovery path, cannot use hidden shortcuts, and becomes stuck in an error state.\\

\midrule

\textbf{Help and Documentation}
& Searching for documentation adds to the agent’s reasoning burden, making the task longer-horizon and more prone to potential cascading errors.
& Navigating to the help page and further retrieve a complete and useful set of information is hard. \\

\bottomrule
\end{tabular}
}
\caption{Five usability heuristics that require adaptation for CUAs, along with the underlying reasoning and representative failure examples. Their original human-centered formulations do not fully address agent-specific constraints in perception, grounding, and control.
}
\label{tab:adaptation_heuristics}
\end{table}

\subsection{Revisiting Nielsen’s ten Usability Heuristics for Interface Design}
\label{sec: revisiting}

Nielsen’s ten usability heuristics provide a compact and influential framework for human-centered interface design, covering principles such as clarity, consistency, error prevention, and user control, with the full list shown in Figure~\ref{fig:examples} (Left). Although these heuristics were not originally developed for computer-use agents, we find that several transfer naturally to agent-compatible interface design, since interface properties that help humans understand interface states and follow effective workflows often also align with agents’ priors and help them as well. In particular, we identify a subset of heuristics that are directly beneficial to CUAs. First, \textbf{consistency and standards} help establish shared structures across interfaces, including common layouts, controls, and descriptions, which support knowledge reuse and improve generalization across applications. Second, \textbf{flexibility and efficiency of use} also matter for agents, as efficient workflows reduce unnecessary interaction steps; this is especially important given that current CUAs still lag behind human experts in execution efficiency~\citep{abhyankar2025osworldhuman, liu2026osexpert}. Third, \textbf{recognition rather than recall} is highly relevant because agents rely on limited context windows to maintain interaction history, making explicit interface cues much easier to use than reconstructing prior states or procedures from memory. Finally, as with humans, \textbf{error prevention} and mechanisms that support \textbf{error diagnosis and recovery} help guide CUAs toward correct workflows and reduce avoidable failures. Taken together, these heuristics can be viewed as directly transferable principles that are broadly beneficial for CUA performance.

While several of Nielsen’s heuristics align naturally with the needs of computer-use agents, others do not transfer as directly. CUAs interact with interfaces under constraints that differ fundamentally from those of human users, including discrete screenshot-based perception, bounded contextual memory, imperfect action grounding, and limited access to implicit shortcuts or recovery strategies. Under these constraints, interface properties that remain intuitive for humans may still introduce ambiguity or execution difficulty for agents. We therefore find that blindly applying some classical heuristics can create unintended barriers for CUA performance. As summarized in Table~\ref{tab:adaptation_heuristics}, these heuristics are not unhelpful in themselves; rather, their original human-centered formulations do not fully account for the perception, grounding, and control requirements of agent interaction. Across these examples, the central issue is that human usability often tolerates implicit cues, metaphor-based semantics, flexible control paths, and documentation-heavy support, whereas CUAs require more explicit, directly executable, and state-aligned interface signals. This observation motivates the augmented heuristics introduced next.

\subsection{General Suggestions for Human-Centered, Agent-Compatible Interface Design}
\label{principle: core}

Interface design can support or hinder reliable CUA automation by shaping how clearly states, actions, and workflows are exposed. Classical usability heuristics may therefore need adaptation for agent operation. Based on this, we offer two general suggestions.

\paragraph{Suggestion 1: Consider agent compatibility early when automation matters.}
When CUA-based automation is intended, agent compatibility should be considered early in the design process. This involves asking whether important states are observable, whether actions are easy to ground, and whether workflows remain stable enough for reliable execution.

\paragraph{Suggestion 2: Keep interface design human-centered while remaining agent-compatible.}
In most applications, CUAs are meant to assist humans rather than fully replace them. Humans should therefore remain able to understand the workflow, supervise agent behavior, and intervene when needed. We advocate a human-centered, agent-compatible design philosophy: interfaces should remain readable and trustworthy for humans while also exposing enough explicit structure for reliable agent interaction.

\subsection{Operationalizing Agent-Compatible Design: CUA-Specific Heuristics}
\label{principle: augment}

Guided by the observations and suggestions above, we propose four CUA-specific usability heuristics that complement classical human-centered interface design. These heuristics are not intended to replace Nielsen’s framework; rather, they serve as structured augmentations for settings where reliable software automation is an explicit goal. Each heuristic highlights interface properties that can reduce exploration burden, improve action grounding, and support more robust agent behavior, even in previously unseen environments.

\begin{augheuristic}{Augmented Interface Usability Heuristic A}
\textbf{Design Computer-Use-Agent-Perceivable Visual States.}
\end{augheuristic}

Interfaces should expose visual states in ways that remain reliably perceivable to CUAs throughout task execution. Critical states and semantics should remain directly visible in the current interface state, even under discrete screen observations, rather than being conveyed through subtle, transient, or difficult-to-detect cues. More broadly, important visual states should always remain explicit, stable, and semantically clear for agent interaction.

\begin{augheuristic}{Augmented Interface Usability Heuristic B}
\textbf{Ensure Stable Layout and Control Consistency.}
\end{augheuristic}

Interfaces should maintain predictable layouts and consistent action semantics across similar contexts. CUAs often struggle when equivalent controls shift position, change appearance, or use inconsistent labels. This may also happen when the control is designed from a physical world metaphor which the agent hardly learned. These scenarios force repeated rediscovery of action mappings. Stable layout and control consistency reduce this burden by allowing learned interaction patterns to transfer across subtasks and interface states. More broadly, this extends classical consistency and standards from semantic consistency to procedural regularity in how controls are organized and exposed for execution.

\begin{augheuristic}{Augmented Interface Usability Heuristic C}
\textbf{Enable Agent-Attainable Step-Wise Control Signals.}
\end{augheuristic}

Interfaces should expose clear and attainable control signals that allow agents to move forward, go back, or recover at each stage of a workflow. For CUAs, control should rely on explicit interface actions rather than hidden shortcuts, implicit gestures, or recovery paths inferred from context. While multiple workflow options may benefit human flexibility, they can also increase ambiguity for agents. A single, explicit control path therefore makes workflows easier for CUAs to follow, verify, and recover within.

\begin{augheuristic}{Augmented Interface Usability Heuristic D}
\textbf{Provide Agent-Friendly Documentation and Explicit Skill Exposure.} 
\end{augheuristic}

Documentation access in digital environments should follow a simple and explicit workflow, imposing minimal grounding, perception, and reasoning burden on agents. More importantly, procedures that agents are expected to use repeatedly should be exposed as explicit, operable skills rather than only as free-form help text. Structured in this way, plain documentation becomes reusable step-by-step procedural knowledge, making the correct operations easier to retrieve, execute, and verify while reducing unnecessary exploration.
\section{Experiments}

\subsection{Experimental Setup}
\label{sec:setup}

\begin{figure*}[t]
    \centering
    \includegraphics[width=1.0\columnwidth]{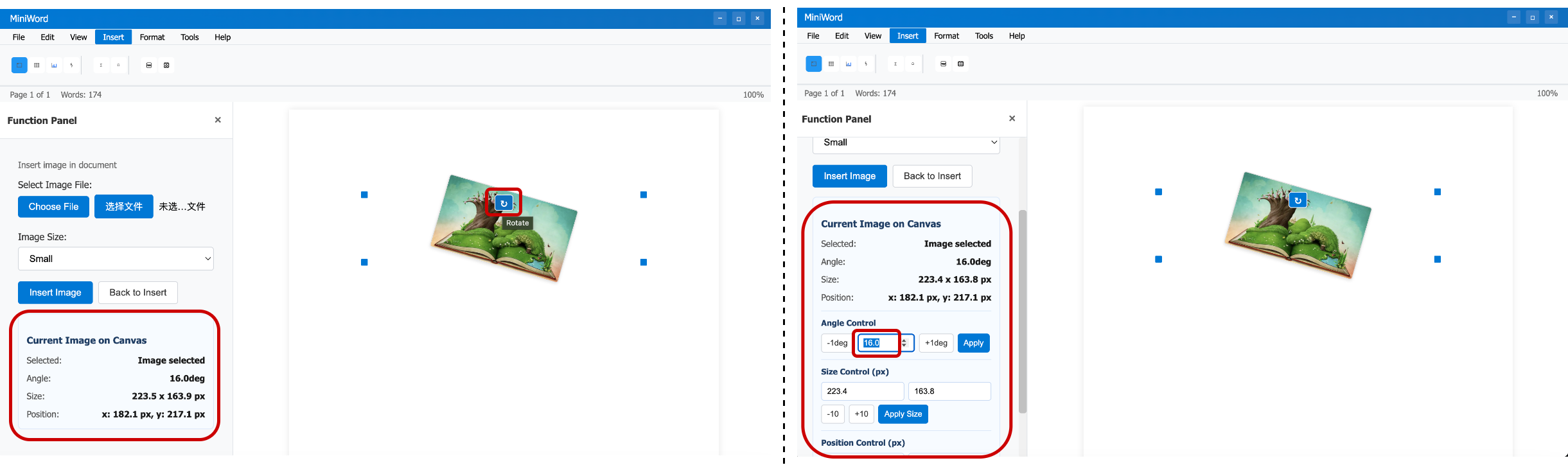}
    \caption{\textbf{Left:} Baseline UI; \textbf{Right:} Revised UI incorporating our augmented usability heuristic B. While the original UI requires direct dragging to mimic rotation or translation, the revised UI adopts a layout that provides explicit controls for rotating and repositioning the target image. More UI examples following our heuristics can be found in Appendix~\ref{app:uiverse}.}
    \label{fig:example_3}
    \vspace{-6pt}
\end{figure*}

We evaluate whether our proposed heuristics for CUAs can improve agent performance without sacrificing human usability. To support this, we introduce UI-Verse (Detailed in Appendix ~\ref{app:uiverse}), a suite of controlled environments built around functionally similar interfaces that support comparable task goals. Each evaluation group contains two interfaces constructed with different sets of usability heuristics. See an example in Figure~\ref{fig:example_3}. We construct baseline UIs that implement Nielsen’s ten usability heuristics while excluding our proposed augmented heuristics. We then build revised UIs that incorporate one or more of our proposed heuristics to better support reliable computer-use agents. For each evaluation group, we manually design computer-use tasks that depend on the target UI regions affected. We then compare agent task success, efficiency, and the distribution of error types in CUA trajectories across the two interface variants.


\paragraph{Model Configurations} For each evaluation group, we test the performance of several types of computer-use agents, including: (1) agents built on general VLMs with UI grounding capability, such as Qwen-3-VL-8B~\cite{bai2025qwen3vltechnicalreport}; (2) agents built on specialized computer-use models fine-tuned on computer-use datasets, such as OpenCUA-7B~\cite{wang2025opencua}; and (3) Agent-S framework~\cite{gonzalezpumariega2025unreasonableeffectivenessscalingagents} that incorporate both a planning module, supported by GPT-5.4-Mini, and a grounding module, based on UI-TARS-7B. To ensure fair evaluation, we do not allow best-of-$N$ sampling at inference time, set the maximum number of interaction steps to 30, and use an inference temperature of 1.0 for all base models. We serve all open-source models locally with vLLM on a single 80GB GPU, and access OpenAI models through the official API. Our environments are implemented on Linux virtual machines and share the same codebase foundation as the widely used computer-use benchmark OSWorld~\citep{xie2024osworld}.

\paragraph{Evaluation Metrics}
We measure agent performance and efficiency using average success rate and average execution time (in seconds) across multiple runs under different evaluation settings and UI design variants. To better understand how the proposed usability heuristics affect agent behavior, we also conduct trajectory-level error attribution. In this stage, a strong LLM serves as an automated evaluator that examines the intermediate interface states and interaction trajectory of each failed run, and classifies the failure as a perception, grounding, or reasoning error. We then report the proportion of each error type among all interaction trajectories to compare failure patterns across evaluation groups.

\begin{table*}[ht]
\centering
\resizebox{1.0\textwidth}{!}
{
    \begin{tabular}{
        c
        l
        |>{\centering\arraybackslash}p{1.8cm}
        |>{\centering\arraybackslash}p{1.8cm}
        |>{\centering\arraybackslash}p{1.8cm}
        |>{\centering\arraybackslash}p{1.8cm}
        |>{\centering\arraybackslash}p{1.8cm}
    }
    \toprule
     \multicolumn{2}{c|}{\textbf{Methods $\rightarrow$}} 
     & \multicolumn{2}{c|}{\textbf{Agent Performance}} 
     & \multicolumn{3}{c}{\textbf{Trajectory Error Attribution}} \\
     \midrule
    \textbf{Models $\downarrow$} & \textbf{Eval $\downarrow$}
    & \textbf{Success} 
    & \textbf{Efficiency} 
    & \textbf{Perception } 
    & \textbf{Grounding } 
    & \textbf{Reasoning } \\
    \midrule

    \rowcolor{gray!15}
    \multicolumn{7}{c}{\textbf{Results w/ and w/o Heuristic A: Agent-Perceivable Visual States}} \\ 
    \midrule
    \multirow{2}{*}{\textbf{\textit{Qwen3-VL}}} 
    & Baseline UI  & 0.13 $\pm$ 0.02  &  102 $\pm$ 11 &0.09 $\pm$ 0.00 & 0.62 $\pm$ 0.00 &0.15 $\pm$ 0.01  \\
    & Revised UI & \textbf{0.31 $\pm$ 0.01} &  \textbf{88 $\pm$ 4} & 0.04 $\pm$ 0.00 & 0.32 $\pm$ 0.01 & 0.33 $\pm$ 0.02 \\
    \midrule
    \multirow{2}{*}{\textbf{\textit{OpenCUA}}} 
    & Baseline UI   & 0.12 $\pm$ 0.00  & 229 $\pm$ 24  & 0.09 $\pm$ 0.00 &0.37 $\pm$ 0.03  & 0.43 $\pm$ 0.03 \\
    & Revised UI & \textbf{0.32 $\pm$ 0.00} &  \textbf{157 $\pm$ 6} & 0.04 $\pm$ 0.00 &0.15 $\pm$ 0.02  & 0.50 $\pm$ 0.02 \\
    \midrule
    \multirow{2}{*}{\textbf{\textit{Agent-S3}}} 
    & Baseline UI   & 0.26 $\pm$ 0.04 &  268 $\pm$ 46 & 0.07 $\pm$ 0.00 & 0.42 $\pm$ 0.02 &0.25 $\pm$ 0.01 \\
    & Revised UI &\textbf{ 0.49 $\pm$ 0.03} &  \textbf{193 $\pm$ 23} & 0.03 $\pm$ 0.00 & 0.12 $\pm$ 0.00 & 0.37 $\pm$ 0.02 \\
    \midrule

    \rowcolor{gray!15}
    \multicolumn{7}{c}{\textbf{Results w/ and w/o Heuristic B: Stable Layout and Control Consistency}} \\ 
    \midrule
    \multirow{2}{*}{\textbf{\textit{Qwen3-VL}}} 
    & Baseline UI   & 0.26 $\pm$ 0.04 &  139 $\pm$ 5 &  0.07 $\pm$ 0.00 & 0.70 $\pm$ 0.05 & 0.09 $\pm$ 0.01 \\
    & Revised UI & \textbf{0.38 $\pm$ 0.03} &  \textbf{138 $\pm$ 2} & 0.06 $\pm$ 0.02 & 0.52 $\pm$ 0.02 & 0.04 $\pm$ 0.00 \\
    \midrule
    \multirow{2}{*}{\textbf{\textit{OpenCUA}}} 
    & Baseline UI   & 0.13 $\pm$ 0.03 & \textbf{ 178 $\pm$ 57} & 0.07 $\pm$ 0.00 & 0.28 $\pm$ 0.02 & 0.53 $\pm$ 0.05 \\
    & Revised UI & \textbf{0.23 $\pm$ 0.03} &  210 $\pm$ 16 & 0.09 $\pm$ 0.00 &0.30 $\pm$ 0.10  & 0.38 $\pm$ 0.08 \\
    \midrule
    \multirow{2}{*}{\textbf{\textit{Agent-S3}}} 
    & Baseline UI   & 0.18 $\pm$ 0.10  &  220 $\pm$ 86 & 0.08 $\pm$ 0.01 & 0.47 $\pm$ 0.03 & 0.28 $\pm$ 0.09 \\
    & Revised UI & \textbf{0.45 $\pm$ 0.21} & \textbf{ 149 $\pm$ 29} &0.08 $\pm$ 0.02  & 0.37 $\pm$ 0.01 &0.18 $\pm$ 0.11  \\
    \midrule

    \rowcolor{gray!15}
    \multicolumn{7}{c}{\textbf{Results w/ and w/o Heuristic C: Enable Step-wise Control Signal}} \\ 
    \midrule
    \multirow{2}{*}{\textbf{\textit{Qwen3-VL}}} 
    & Baseline UI   & 0.10 $\pm$ 0.00 & 121 $\pm$ 7 & 0.09 $\pm$ 0.01 & 0.71 $\pm$ 0.05 & 0.10 $\pm$ 0.04 \\
    & Revised UI & \textbf{0.30 $\pm$ 0.00} & \textbf{ 64 $\pm$ 4} & 0.07 $\pm$ 0.00 & 0.51 $\pm$ 0.00 & 0.12 $\pm$ 0.00 \\
    \midrule
    \multirow{2}{*}{\textbf{\textit{OpenCUA}}} 
    & Baseline UI   & 0.15 $\pm$ 0.07 & \textbf{224 $\pm$ 8 }& 0.08 $\pm$ 0.01 & 0.26 $\pm$ 0.07 & 0.52 $\pm$ 0.03 \\
    & Revised UI & \textbf{0.35 $\pm$ 0.07} &  243 $\pm$ 28 & 0.05 $\pm$ 0.01 & 0.32 $\pm$ 0.05 &  0.28 $\pm$ 0.09\\
    \midrule
    \multirow{2}{*}{\textbf{\textit{Agent-S3}}} 
    & Baseline UI  & 0.20 $\pm$ 0.00 & \textbf{ 212 $\pm$ 5} & 0.08 $\pm$ 0.01 & 0.40 $\pm$ 0.00 & 0.33 $\pm$ 0.01 \\
    & Revised UI & \textbf{0.40 $\pm$ 0.14 }& 224 $\pm$ 24  & 0.05 $\pm$ 0.02 &  0.38 $\pm$ 0.10&  0.17 $\pm$ 0.02\\
    \midrule

    \rowcolor{gray!15}
    \multicolumn{7}{c}{\textbf{Results w/ and w/o Heuristic D: Agent-friendly Documentation and Skill Exposure}} \\ 
    \midrule
    \multirow{2}{*}{\textbf{\textit{Qwen3-VL}}} 
    & Baseline UI   & 0.03 $\pm$ 0.00 &  155 $\pm$ 3 & 0.10 $\pm$ 0.00 &0.70 $\pm$ 0.01  & 0.18 $\pm$ 0.01 \\
    & Revised UI & \textbf{0.38 $\pm$ 0.00} &  \textbf{86 $\pm$ 7} & 0.08 $\pm$ 0.00 & 0.47 $\pm$ 0.01 & 0.07 $\pm$ 0.01 \\
    \midrule
    \multirow{2}{*}{\textbf{\textit{OpenCUA}}} 
    & Baseline UI   & 0.08 $\pm$ 0.04 &  264 $\pm$ 1 & 0.08 $\pm$ 0.00 &0.37 $\pm$ 0.01  & 0.46 $\pm$ 0.01 \\
    & Revised UI & \textbf{0.23 $\pm$ 0.00} & \textbf{ 207 $\pm$ 1} & 0.11 $\pm$ 0.00 &0.37 $\pm$ 0.03  & 0.29 $\pm$ 0.03 \\
    \midrule
    \multirow{2}{*}{\textbf{\textit{Agent-S3}}} 
    & Baseline UI   & 0.18 $\pm$ 0.04 &  230 $\pm$ 84 & 0.08 $\pm$ 0.01 &0.44 $\pm$ 0.02  & 0.30 $\pm$ 0.00 \\
    & Revised UI & \textbf{0.41 $\pm$ 0.04} & \textbf{ 195 $\pm$ 17} & 0.08 $\pm$ 0.00 & 0.39 $\pm$ 0.06 & 0.12 $\pm$ 0.04 \\
    \midrule

    \rowcolor{gray!15}
    \multicolumn{7}{c}{\textbf{Results w/ and w/o All Augmented Heuristics A-D}} \\ 
    \midrule
    \multirow{2}{*}{\textbf{\textit{Qwen3-VL}}} 
    & Baseline UI   & 0.13 $\pm$ 0.02 & 129 $\pm$ 7 & 0.09 $\pm$ 0.00 & 0.68 $\pm$ 0.03 & 0.10 $\pm$ 0.02 \\
& Revised UI    & \textbf{0.59 $\pm$ 0.03} & \textbf{62 $\pm$ 5} & 0.05 $\pm$ 0.00 & 0.28 $\pm$ 0.02 & 0.08 $\pm$ 0.02 \\
\midrule
\multirow{2}{*}{\textbf{\textit{OpenCUA}}} 
& Baseline UI   & 0.12 $\pm$ 0.04 & 224 $\pm$ 23 & 0.08 $\pm$ 0.00 & 0.32 $\pm$ 0.03 & 0.48 $\pm$ 0.03 \\
& Revised UI    & \textbf{0.54 $\pm$ 0.04} & \textbf{149 $\pm$ 11} & 0.04 $\pm$ 0.00 & 0.20 $\pm$ 0.02 & 0.22 $\pm$ 0.03 \\
\midrule
\multirow{2}{*}{\textbf{\textit{Agent-S3}}} 
& Baseline UI   & 0.21 $\pm$ 0.05 & 280 $\pm$ 55 & 0.08 $\pm$ 0.01 & 0.43 $\pm$ 0.02 & 0.28 $\pm$ 0.03 \\
& Revised UI    & \textbf{0.61 $\pm$ 0.04} & \textbf{140 $\pm$ 16} & 0.05 $\pm$ 0.00 & 0.20 $\pm$ 0.02 & 0.14 $\pm$ 0.03 \\
    \bottomrule
    \end{tabular}
}
\caption{Agent performance and trajectory error attribution on baseline UIs designed under Nielsen’s 10 usability heuristics and revised UIs augmented with our proposed agent-compatible heuristics. Across settings, implementing any of the proposed heuristics generally improves task success and reduces specific failure types, highlighting how different interface designs shape robust CUA behavior in perception, grounding, and reasoning.}
\label{tab: main results}
\end{table*}

\subsection{Main Results}
\label{sec:main results}

\paragraph{Consistent performance gains from pure UI design shifts.}
The left sections of Table~\ref{tab: main results} presents our main finding: improving agent compatibility through UI design alone can substantially improve CUA performance on selected tasks. Across models and agentic frameworks, applying any single one of our proposed heuristics consistently increases task success rate relative to the baseline UI. The largest absolute gain reaches 35\% for Qwen3-VL under Heuristic D, showing that improvement can be substantial. In addition to higher success rates, we also observe  efficiency gains in most settings. A likely explanation is that improved interface design helps agents complete tasks more directly, rather than becoming trapped in error states and repeatedly exploring until the maximum interaction budget is exhausted. Overall, these results suggest that UI design itself is a strong and underexplored lever for improving the robustness and efficiency of computer-use agents, and that our proposed heuristics are practically useful for supporting more reliable agent interaction.

\paragraph{Joint and complementary benefits from combining all proposed heuristics.}

As shown in Table~\ref{tab: main results}, when all four augmented heuristics are implemented together, all evaluated agents achieve substantial improvements over the baseline UI. For example, Qwen3-VL improves from 0.13 to 0.59, corresponding to an absolute gain of 46\% and roughly a 3.5$\times$ improvement over the baseline, while Agent-S3 improves from 0.21 to 0.61, corresponding to an absolute gain of 40\%. These joint gains are also larger than those achieved under any single heuristic alone, indicating that the four heuristics provide complementary benefits when combined. They are further accompanied by large reductions in execution time, as well as clear decreases in grounding and reasoning errors, with perception errors also decreasing more modestly. Overall, this pattern suggests that the heuristics target different failure modes in agent interaction and therefore work better together than in isolation.


\paragraph{Error-type shifts under different heuristic implementations.}
The right half of Table~\ref{tab: main results} reports trajectory-level error attribution under the baseline and revised UIs. Different heuristics reduce different failure types, in ways that largely match their intended roles. Heuristic A most consistently improves grounding, with smaller gains in perception, suggesting that clearer visual states help agents recognize interface state and ground actions correctly. Heuristic B and Heuristic C mainly reduce reasoning errors, indicating that stable layouts, consistent controls, and step-wise interaction signals help agents plan, follow, and recover within workflows. Heuristic D yields the strongest reduction in reasoning errors, while also improving grounding in several cases, suggesting that structured documentation and explicit skill exposure reduce procedural burden.

\subsection{Human Studies}

\begin{figure*}[t]
    \centering
    \includegraphics[width=1.0\columnwidth]{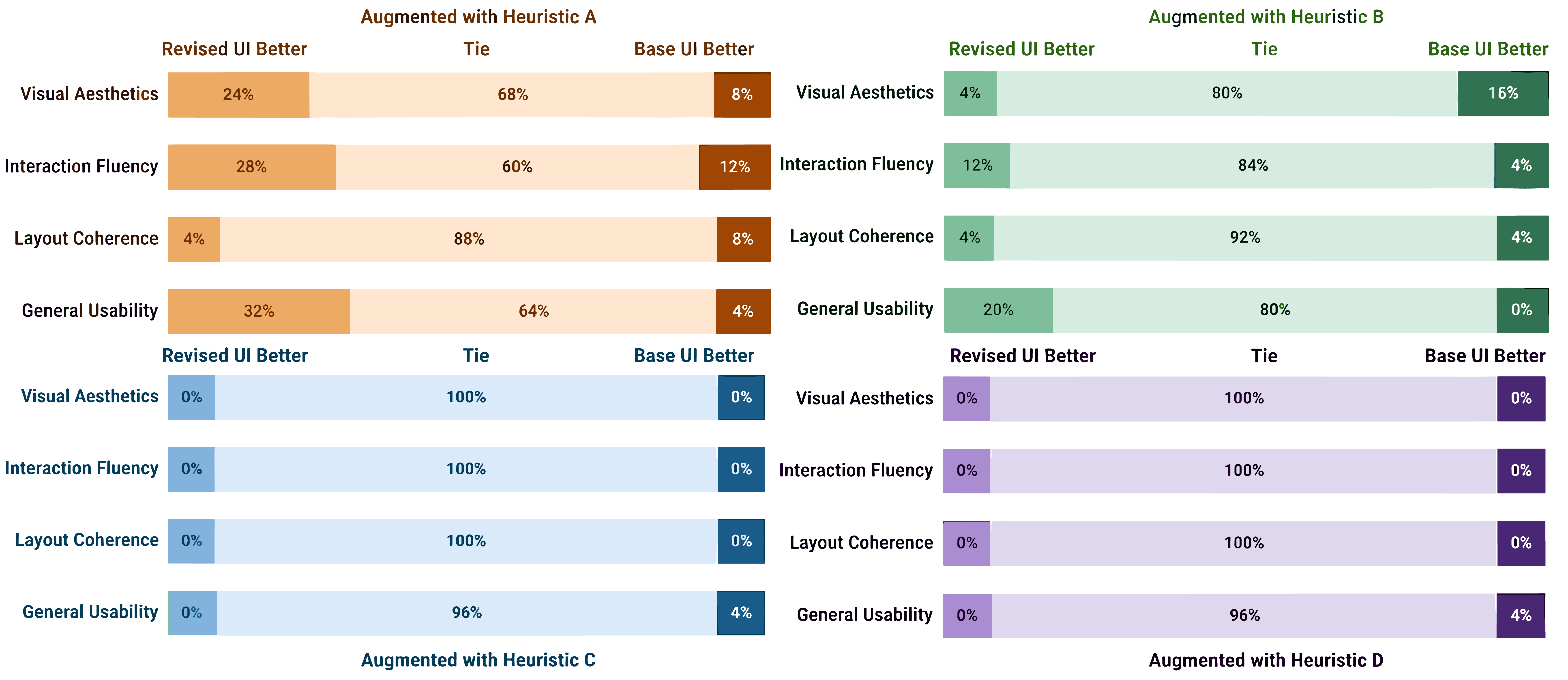}
    \caption{Human evaluation of revised interfaces under heuristic augmentations. Across all four revised UIs, annotators overwhelmingly judge the revised and base interfaces as comparable on visual aesthetics, interaction fluency, layout coherence, and general usability. This indicates that our agent-oriented revisions improve CUA compatibility without materially degrading human usability.}
    \label{fig:human}
    \vspace{-6pt}
\end{figure*}

We conduct comprehensive human studies with two goals: (1) to evaluate whether our agent-compatible heuristics preserve human usability, and (2) to assess the reliability of our automatic trajectory error analyzer. Specifically, we examine whether the revised interfaces remain effective for human users, while also measuring how closely the analyzer’s judgments align with human annotations on error attribution. Additional details on annotators, experimental configurations, and more comprehensive results are provided in Appendix~\ref{app:exp}.

\paragraph{Human-Centered Evaluation of Revised Interfaces}
In line with our general suggestion in Section~\S\ref{principle: core}, we conduct human studies to examine whether the proposed heuristics preserve interface quality for human users, rather than improving agent performance at the expense of human experience. Annotators compare randomly sampled screenshots and standard interaction trajectories from selected tasks under both the baseline and revised interfaces along four dimensions: \textbf{visual aesthetics}, \textbf{interaction fluency}, \textbf{layout coherence}, and \textbf{perceived overall usability}. For each task, they are shown six screenshots from the baseline interface and six corresponding screenshots from the revised interface, and judge whether the revised UI is better, the baseline UI is better, or whether the two are tied on each dimension. As shown in Figure~\ref{fig:human}, most judgments are ties across all four heuristic-based revisions, indicating that the augmented designs generally preserve human usability. In some cases, especially under Heuristic A, annotators slightly prefer the revised UI. Overall, these results suggest that our augmented heuristics improve support for CUAs without meaningfully harming human experience.

\paragraph{Reliability of Automatic Error Analysis}
We further evaluate the reliability of our automatic trajectory error analyzer by comparing its judgments with human annotations. Human evaluators inspect failed interaction trajectories and estimate the extent to which each failure is primarily caused by perception, grounding, or reasoning errors. As shown in Table~\ref{tab:error_human_eval} (Appendix~\ref{app:exp}), the automatic analyzer produces error categorizations that are broadly consistent with human judgments across all four heuristics. In particular, both human and automatic evaluations reveal similar distribution shifts in how revised interfaces reduce different types of agent errors, supporting our core findings in Section~\ref{sec:main results}.

\section{Conclusion}

In this paper, we argue that improving computer-use agents should not be treated solely as a model-side problem, but also as a problem of agent-compatible interface design. From this perspective, we revisit Nielsen’s usability heuristics through the lens of CUAs, identify where classical human-centered design transfers and where adaptation is needed, and propose a human-centered, agent-compatible framework with two general design suggestions and four concrete heuristics. To study this question, we introduce UI-Verse, a controlled benchmark for analyzing how interface design affects agent behavior, and evaluate multiple CUAs on interfaces with and without our augmented heuristics. Results show that agent-compatible interfaces can consistently improve task success and often efficiency, while human studies suggest these gains can be achieved without clear usability regressions. Overall, we view this work as an initial step toward co-designed digital environments that are usable for humans and dependable for general CUAs, enabling more robust automation and more effective human--AI collaboration in both everyday and professional workflows.

\section*{Reproducibility Statement}

In Section~\S\ref{sec:setup}, we describe our experimental setup in detail, including the model configurations, computational resource requirements, and access to the models used in our study. We also report the core implementation details needed to reproduce our experiments. To promote transparency and enable future extensions of this work, we will publicly release the codebase and data upon acceptance. These resources are intended to support the replication of our main results and provide a foundation for subsequent research. 


\bibliography{colm2026_conference}
\bibliographystyle{colm2026_conference}

\newpage
\appendix

\section{UI-Verse with Augmented Usability Heuristics}
\label{app:uiverse}

In this section, we introduce UI-Verse, a suite of controlled environments built around functionally similar interfaces that support comparable task goals. Within each evaluation group, UI-Verse includes two interfaces constructed with different sets of usability heuristics. By evaluating CUAs on the same tasks across these interface variants, we can quantitatively examine how interface usability heuristics affect agent performance. We provide the list of implemented unit functionalities for environments in UI-Verse in Section~\ref{sec:lof}, and present exemplar UI designs in Section~\ref{sec:examplar} to illustrate how our proposed heuristics may map to real-world interface design choices.

\subsection{List of Functionalities Implemented}
\label{sec:lof}

In our experiments, we implement 63 unit functionalities for document processing in MiniWord. 
All UI-Verse environments preserve the same underlying functionality set, while allowing the interface realization to vary according to different usability heuristics. 
This design ensures that performance differences arise from interface design rather than functional disparity. 
Table~\ref{tab:miniword_functionality_grid} summarizes all implemented unit functionalities. 
We leave the extension toward broader and more complex functionality coverage to future work.

\begin{table*}[t]
\centering
\footnotesize
\renewcommand{\arraystretch}{1.2}
\setlength{\tabcolsep}{3pt}
\resizebox{\textwidth}{!}{%
\begin{tabular}{|p{2.0cm}|p{2.0cm}|p{2.0cm}|p{2.0cm}|p{2.0cm}|p{2.0cm}|p{2.0cm}|}
\hline
\rowcolor{blue!18}
\multicolumn{7}{|c|}{\textbf{All functionalities imeplmented in ENvironements of UI-Verse}} \\
\hline
\rowcolor{blue!8}
New Doc & Open Doc & Save Doc & Print Doc & Refresh Page & Close Page & New Tab \\
\hline
\rowcolor{gray!12}
Pagination & Undo & Redo & Cut & Direct Copy & Format Copy & Paste \\
\hline
\rowcolor{blue!8}
Find Text & Replace & Clear Format & Painter & Zoom In & Zoom Out & Print Preview \\
\hline
\rowcolor{gray!12}
Page Setup & Toggle Bar & Toggle Ruler & Hide & Focus Mode & Night Mode & Insert Image \\
\hline
\rowcolor{blue!8}
Insert Table & Insert Chart & Insert Link & Equation & Insert Symbol & Section Break & Use Template \\
\hline
\rowcolor{gray!12}
Bold & Italic & Underline & Strikethrough & Left Align & Center Align & Right Align \\
\hline
\rowcolor{blue!8}
Justify & Text Color & Highlight & Format Paint & Clear Format & Superscript & Subscript \\
\hline
\rowcolor{gray!12}
Line Spacing & Page Setup & Header Footer & Spell Check & Comment & Bulleted List & Numbere List \\
\hline
\rowcolor{blue!8}
Add Indent & Cut Indent & Count & Settings & Page Break & Share Doc & Help Doc \\
\hline
\end{tabular}%
}
\caption{All implemented unit functionalities in MiniWord. Each cell corresponds to one functionality shared across all UI-Verse environments. While interface layouts may vary under different usability heuristics, the underlying functionality set is kept fixed.}
\label{tab:miniword_functionality_grid}
\end{table*}

\begin{figure*}[t]
    \centering
    \includegraphics[width=1.0\columnwidth]{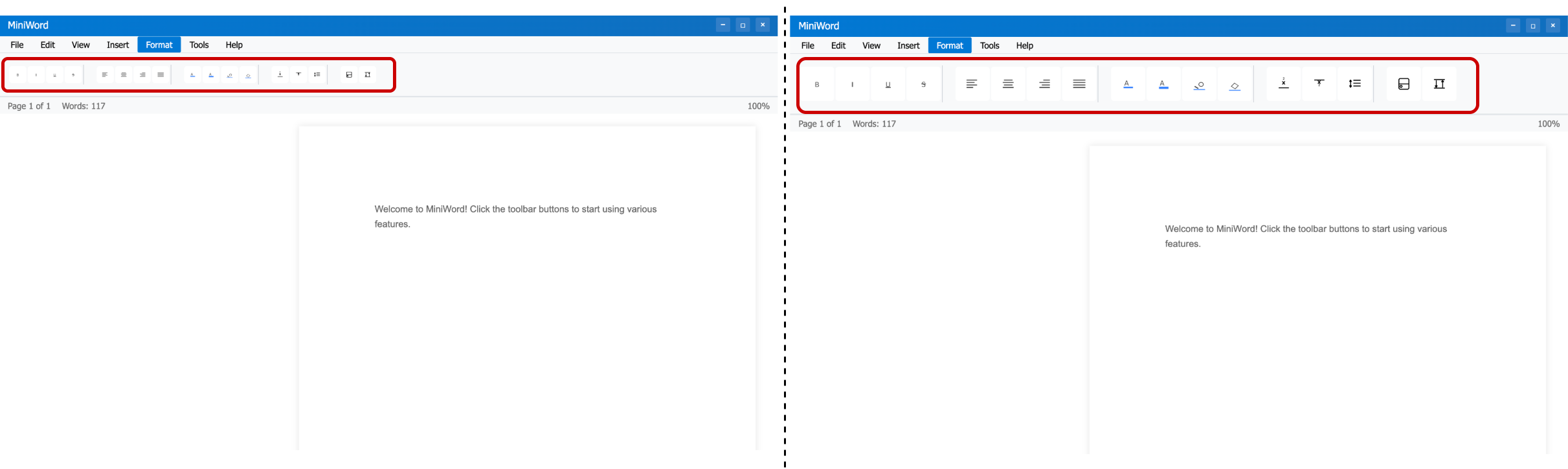}
    \caption{\textbf{Left:} Baseline UI; \textbf{Right:} Revised UI incorporating  our augmented usability heuristics A. Specifically, the revised UI only uses larger buttons for better agent visibility.}
    \label{fig:example_1}
    \vspace{-6pt}
\end{figure*}

\begin{figure*}[t]
    \centering
    \includegraphics[width=1.0\columnwidth]{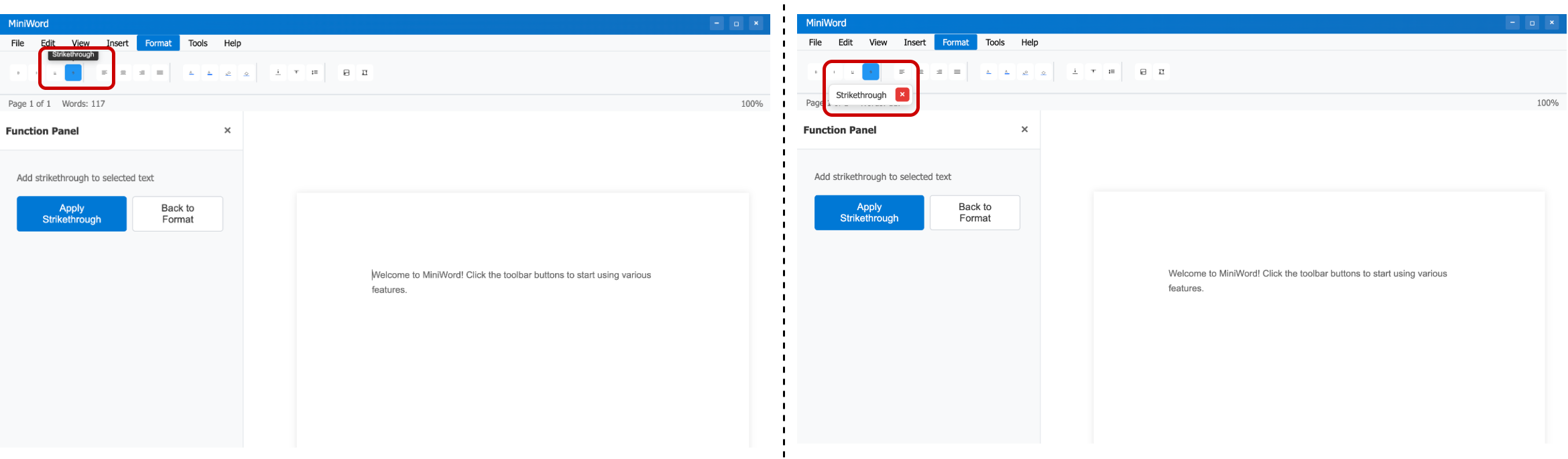}
    \caption{\textbf{Left:} Baseline UI; \textbf{Right:} Revised UI incorporating our augmented usability heuristic A. Specifically, the revised UI avoids hover-only textual descriptions of buttons, making functionality labels visible and reducing agent confusion about button meanings.}
    \label{fig:example_2}
    \vspace{-6pt}
\end{figure*}

\begin{figure*}[t]
    \centering
    \includegraphics[width=1.0\columnwidth]{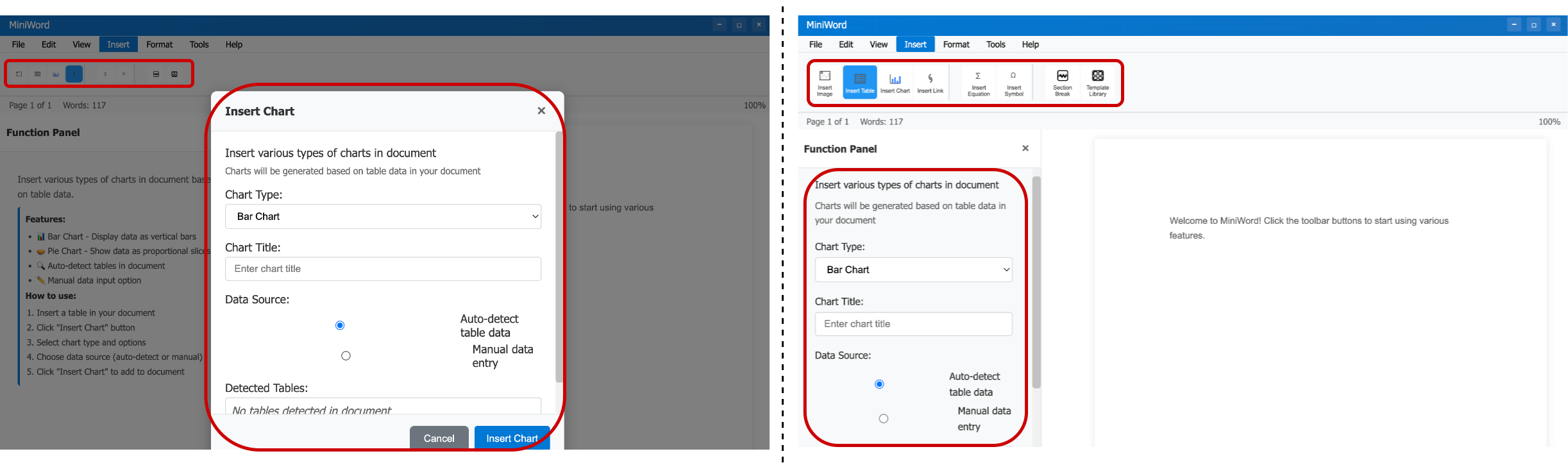}
    \caption{\textbf{Left:} Baseline UI; \textbf{Right:} Revised UI incorporating all of our augmented usability heuristics for CUAs. Specifically, the revised UI uses larger buttons, adds text-based functionality descriptions to improve agent visibility, and adopts a more consistent layout by always using sidebars. Beyond the visible interface changes, the revised environment also provides more consistent controls, agent-executable shortcuts for directly returning to the previous state, and a documented skill set that can be leveraged by CUAs.}
    \label{fig:example_4}
    \vspace{-6pt}
\end{figure*}

\subsection{Augmented Heuristics in Interface Design}
\label{sec:examplar}

Although each of our augmented heuristics (A--D) can be implemented in multiple ways, we provide representative examples to illustrate how baseline interfaces can be revised to better support agent interaction. Figures~\ref{fig:example_1}, \ref{fig:example_2}, and \ref{fig:example_4} present example pairs of baseline and revised UI designs from our UI-Verse environments. The key interface changes are highlighted with red boxes in the figures, while the individual captions further explain the specific modifications and why the revised designs are more helpful for agents.

\section{Additional Results for Human Studies}
\label{app:exp}

\begin{table*}[t]
\centering
\small
\setlength{\tabcolsep}{4pt}
\renewcommand{\arraystretch}{1.15}
\resizebox{\textwidth}{!}{%
\begin{tabular}{l|cc cc cc|cc cc cc}
\toprule
\textbf{Augmentation $\rightarrow$} 
& \multicolumn{6}{c|}{\textbf{Heuristic A}} 
& \multicolumn{6}{c}{\textbf{Heuristic B}} \\
\cline{2-7} \cline{8-13}
\textbf{Error Type $\rightarrow$} 
& \multicolumn{2}{c}{\textbf{Perception}} 
& \multicolumn{2}{c}{\textbf{Grounding}} 
& \multicolumn{2}{c|}{\textbf{Reasoning}} 
& \multicolumn{2}{c}{\textbf{Perception}} 
& \multicolumn{2}{c}{\textbf{Grounding}} 
& \multicolumn{2}{c}{\textbf{Reasoning}} \\
\cline{2-3} \cline{4-5} \cline{6-7}
\cline{8-9} \cline{10-11} \cline{12-13}
\textbf{Eval $\downarrow$} 
& \textbf{Auto} & \textbf{Human} 
& \textbf{Auto} & \textbf{Human} 
& \textbf{Auto} & \textbf{Human} 
& \textbf{Auto} & \textbf{Human} 
& \textbf{Auto} & \textbf{Human} 
& \textbf{Auto} & \textbf{Human} \\
\midrule
Baseline UI 
& 0.07 & 0.15
& 0.42 & 0.33
& 0.25 & 0.26
& 0.08 & 0.14
& 0.47 & 0.36
& 0.28 & 0.33 \\
Revised UI  
& 0.03 & 0.08
& 0.12 & 0.11
& 0.37 & 0.33
& 0.08 & 0.11
& 0.37 & 0.27
& 0.18 & 0.25 \\
\midrule
\midrule
\textbf{Augmentation $\rightarrow$} 
& \multicolumn{6}{c|}{\textbf{Heuristic C}} 
& \multicolumn{6}{c}{\textbf{Heuristic D}} \\
\cline{2-7} \cline{8-13}
\textbf{Error Type $\rightarrow$} 
& \multicolumn{2}{c}{\textbf{Perception}} 
& \multicolumn{2}{c}{\textbf{Grounding}} 
& \multicolumn{2}{c|}{\textbf{Reasoning}} 
& \multicolumn{2}{c}{\textbf{Perception}} 
& \multicolumn{2}{c}{\textbf{Grounding}} 
& \multicolumn{2}{c}{\textbf{Reasoning}} \\
\cline{2-3} \cline{4-5} \cline{6-7}
\cline{8-9} \cline{10-11} \cline{12-13}
\textbf{Eval $\downarrow$} 
& \textbf{Auto} & \textbf{Human} 
& \textbf{Auto} & \textbf{Human} 
& \textbf{Auto} & \textbf{Human} 
& \textbf{Auto} & \textbf{Human} 
& \textbf{Auto} & \textbf{Human} 
& \textbf{Auto} & \textbf{Human} \\
\midrule
Baseline UI 
& 0.08 & 0.12
& 0.44 & 0.38
& 0.30 & 0.32
& 0.08 & 0.13
& 0.43 & 0.34
& 0.28 & 0.32 \\
Revised UI  
& 0.08 & 0.10
& 0.39 & 0.29
& 0.12 & 0.20
& 0.05 & 0.09
& 0.20 & 0.16
& 0.14 & 0.14 \\
\bottomrule
\end{tabular}%
}
\caption{Human validation of automatic trajectory error attribution under different interface heuristics, based on trajectories generated by the Agent-S framework. While the automatic evaluator and human annotators do not always produce identical absolute error proportions, they exhibit strong agreement in overall error patterns. Notably, both capture similar shifts in the distribution of perception, grounding, and reasoning errors introduced by different interface heuristics.}
\label{tab:error_human_eval}
\end{table*}

\begin{figure*}[t]
    \centering
    \includegraphics[width=1.0\columnwidth]{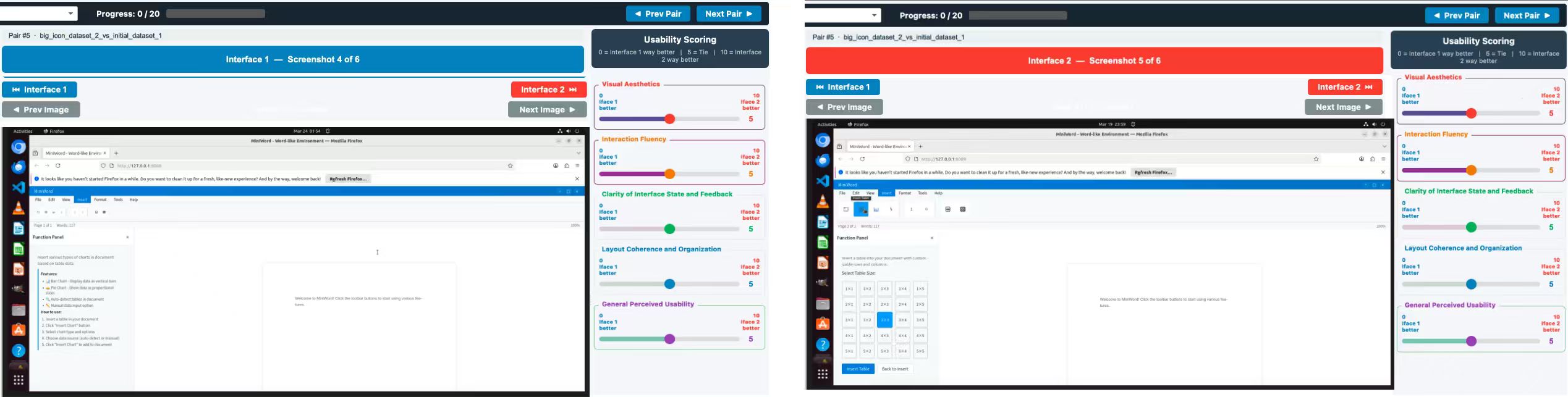}
    \caption{Human annotation interface used in the first part of our human study.}
    \label{fig:annotation_ui_1}
    \vspace{-6pt}
\end{figure*}

\begin{figure*}[t]
    \centering
    \includegraphics[width=1.0\columnwidth]{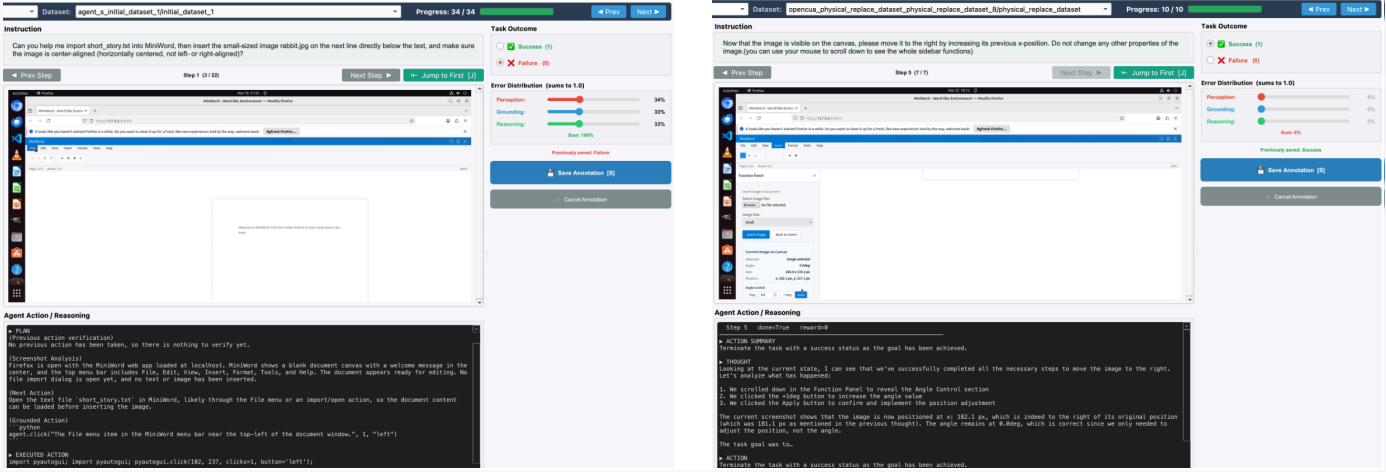}
    \caption{Human annotation interface used in the second part of our human study.}
    \label{fig:annotation_ui_2}
    \vspace{-6pt}
\end{figure*}

\subsection{Experimental Settings}
Four volunteer researchers with backgrounds in computer science and AI participated in our human studies. We designed separate annotation interfaces for the two experiments, as shown in Figure~\ref{fig:annotation_ui_1} and Figure~\ref{fig:annotation_ui_2}.

In the first study, annotators were first introduced to the definitions of the evaluation dimensions for human-centered UI quality. They were then presented with six screenshots from the baseline interface and six corresponding screenshots from the revised interface, and asked to compare the two interfaces and judge which one better matched their expectations along each dimension. The annotation interface supports efficient switching across screenshots and provides shortcuts for quickly toggling between the two interfaces.

In the second study, annotators were first introduced to the definitions of erroneous trajectories and the three error categories. They then examined the interaction trajectories between CUAs and the interfaces, including the agent’s intermediate planning, reasoning, actions, and the resulting UI changes over time. Based on the full trajectory, annotators used the interface to assign a probability distribution over the three error types: perception, grounding, and reasoning. The annotation interface supports efficient navigation across screenshots and provides shortcuts for jumping directly to the final stage. The agent’s reasoning and thoughts are presented in natural language in a terminal panel. 

\subsection{More Detailed Experimental Results}
Due to space limitations, we provide the detailed results for the second part of the human study in Table~\ref{tab:error_human_eval}. These results show that although human judgments and the automatic analyzer may differ slightly in absolute values, they remain broadly consistent overall. In particular, both exhibit similar distribution shifts under our usability heuristics, further supporting the reliability of the automatic analysis.

\section{Discussions}

\paragraph{Scope and boundary of our setting.}
Our work focuses on a particular and practically important stream of computer-use agents: general, UI-centered agents that operate existing software through the same interface surface shared with human users. This framing is aligned with our paper’s core setting, where interfaces are treated as a shared medium for both humans and agents, with an emphasis on preserving human supervision, interpretability, and intervention when needed. In this sense, our study is most relevant to workflows in which software remains human-facing and agents must generalize through the visible interface, rather than through privileged back-end access.

\paragraph{Automation-native software beyond graphical interfaces.}
We acknowledge that UI-based interaction represents only one stream within the broader agentic software landscape. An important alternative direction is automation-native software, in which systems are designed from the outset for direct machine interaction through APIs, protocols, or structured tool interfaces rather than human-facing graphical UIs. Recent developments point toward this paradigm. The Model Context Protocol (MCP), for example, standardizes how AI systems connect to external tools and data sources, enabling more direct and structured integration beyond GUI operation \citep{mcp_official_2025,hou2025mcp,hasan2025mcpfirstglance}. Similarly, CLI-Anything advocates making software directly agent-native through command-line-style interfaces, thereby reducing reliance on fragile screen-based interaction \citep{clianything_2026}. Our paper is not primarily targeted at these non-UI-first workflows. Instead, we focus on the complementary regime in which agents must still operate through existing graphical interfaces shared with human users. More broadly, we believe that even as AI agents become increasingly capable, human supervision will remain important in many real-world workflows, especially when trustworthiness, safety, and accountability are critical. In such settings, interfaces that are understandable to human users remain necessary not only to support oversight and intervention, but also to reduce cognitive burden and make collaboration with agents more practical.

\end{document}